 \documentstyle[12pt]{article}
\def\prb{Phys. Rev. B }
\def\prl{Phys. Rev. Lett. }

\begin{document}
\baselineskip .3in
\pagestyle{empty}
\sloppy
\newpage
\pagestyle{plain}

\title{Amplification or reduction of backscattering in a coherently 
amplifying or absorbing disordered chain}

\author{Asok K. Sen$^*$ \\
{\it International Centre for Theoretical Physics}\\
{\it P.O. Box 586, 34100 Trieste, Italy}\\
and \\
{\it LTP Division, Saha Institute of Nuclear Physics}\\
{\it 1/AF, Bidhannagar, Calcutta 700 064, India}}

\date{\today}
\maketitle

\begin{abstract}
We study localization properties of a one-dimensional disordered system
characterized by a random non-hermitean hamiltonian where both the
randomness and the non-hermiticity arises in the local site-potential;
its real part being random, and a constant imaginary part implying the
presence of either a coherent absorption or amplification at each site. 
While the two-probe transport properties behave seemingly very
differently for the amplifying and the absorbing chains, the logarithmic
resistance $u$ = ln$(1+R_4)$ where $R_4$ is the 4-probe resistance gives
a unified description of both the cases.  It is found that the
ensemble-averaged $<u>$ increases linearly with length indicating exponential
growth of resistance.  While in contrast to the case of Anderson localization
(random hermitean matrix), the variance of $u$ could be orders of magnitude
smaller in the non-hermitean case, the distribution of $u$ still remains
non-Gaussian even in the large length limit. 
\end{abstract}

{PACS numbers: 05.40.+j, 42.25.Bs, 71.55.Jv, 72.15.Rn}

$^*${\it e-mail address: asok@hp2.saha.ernet.in}

\newpage

The study of interference of waves multiply scattered from a system of
scatterers with non-hermitean hamiltonians has of late become very 
fashionable.  There are two classes of problems in this respect; one
in which the non-hermiticity is in the nonlocal part \cite{nlnh1, nlnh2}
of the hamiltonuan and the other in which the non-hermiticity is
in the local part (typically in one-body potentials) \cite
{john,expt,weav,kum,frei1,gj,mpl1,pass,zhn,been,frei2,misi,yose,frei3}.
The first category is understood to represent, among other things,
the physics of vortex lines pinned by columnar defects where
the depinning is achieved \cite{nlnh1} by a sufficiently high transverse
magnetic field (represented by an imaginary vector potential).  In the second
category, it is well-known that an imaginary term in the local part of the
non-hermitean hamiltonian behaves like a source or a sink (depending on the
sign).  We will be concerned here with non-hermiticity in the one-body
potential only.  In a disordered chain with random but real-valued
site-potentials, almost all the states are exponentially localized and hence
an incident wave ($\sim e^{ikx}$) propagating in the positive $x$-direction
is completely backscattered due to the well-known localization effects
\cite{ramps}.  On the other hand, a purely ordered chain with fixed
absorbing site-potentials (sinks for particles) is intuitively expected to
lead to exponential decay of the transmittance (forward-scattering), and so
it does.  While similar intuition may lead to the expectation that the
transmittance would increase indefinitely if each of the fixed imaginary
site-potentials is amplifying (source of particles), in actuality it does not.
In a recent paper \cite{mpl1} (referred to as I from now on), it was shown
by the author that the sample with amplifying
sites do also behave as a perfect reflector in the large length limit,
because the perfect lead boundary conditions at the two ends of the chain
imposes a real energy spectrum for the non-hermitean matrix \cite{f1}.
In this work, we generalize over our work in I and consider the effects of
disorder on the coherent amplification/ absorption at each site
and look for an unified description for both the cases.

We consider a quantum chain of $N$ lattice points (lattice constant unity),
represented by the standard single band, tight binding equation:
\begin{equation}
      (E - \epsilon_n) c_n = V(c_{n-1} + c_{n+1}).
\end{equation}
This {\it open quantum system} is coupled to the external world (two
reservoirs at very slightly different electrochemical potentials) with two
identical semi-infinite perfect leads on either side.  Here $E$ is the fermionic
energy, $V$ is the constant nearest neighbor hopping term which is the same
in both the leads and the sample, $\epsilon_n$ is the site-energy, and $c_n$
is the site amplitude at the $n$th site.  Without any loss of generality, we
choose $\epsilon_n = 0$ in the leads and $V=1$ to set the energy scale.  Inside
the sample, we choose $\epsilon_n = \epsilon_r + i\eta$ where the real part
$\epsilon_r$ is random and the imaginary part $\eta$ is a fixed real number
which may be either positive or negative, and $i=\sqrt{-1}$.  The random
$\epsilon_r$ is obtained from an uniform distribution on $[-W/2,W/2]$.  Since for
an isolated scattering potential with a positive (negative) $\eta$ the wave-vector
($k$) has a positive (negative) imaginary part, the wave ($\sim e^{ikx}$) decays
(grows) exponentially with $x$. Thus a medium with positive $\eta$ at each site
is called an absorbing medium and a medium with negative $\eta$ an amplifying
medium. The physical reason for such a description lies in the fact that the
scattering in any real medium is never perfectly elastic and that in many cases
the deviation from perfectly elastic scattering may be described by absorption
through other inelastic channels or amplification due to enhancement of the
wave-amplitude (e.g., population inversion in an active medium) of incident
particles or waves.
The complex transmission amplitude in the ordered case was calculated to be
\begin{equation}
t_A={(e^{ik_s}e^{-\gamma}-e^{-ik_s}e^{\gamma})(e^{ik}-e^{-ik})e^{-ik(L+2)}
\over de^{-ik_sL}e^{\gamma L} - ce^{ik_sL}e^{-\gamma L}},
\end{equation}
\noindent where
\begin{equation}
      c=(e^{ik_s-ik}e^{-\gamma} - 1)^2,
      d=(e^{-ik_s-ik}e^{\gamma} - 1)^2,
\end{equation}
\noindent and the decay length $1/|\gamma|=l_a$ and the wave-vector $k_s$ are
given by
\begin{equation}
	E = 2{\rm cos} k = (e^{\gamma} + e^{-\gamma}){\rm cos} k_s ,
\end{equation}
and
\begin{equation}
	 \eta = (e^{\gamma} - e^{-\gamma}){\rm sin} k_s .
\end{equation}
The transmittance or the two-probe conductance $T=g_2=|t_A|^2$ obtained from
the Eq.(2) is found to decay monotonically (exponentially) towards zero for a
set of absorbers ($\eta > 0$).  But, for a set of amplifiers ($\eta < 0$), 
$g_2$ increases first to a high value but eventually (for large $L$) decays as
$t_A \sim e^{-|\gamma|L}$.  Disorder (in $\epsilon_r$) in 1D is known to give
rise to an exponential decay.  Thus in the presence of both disorder and
absorption/ amplification, one expects the conductance to behave for $L \to
\infty$ as
\begin{equation}
	t \sim t_Ae^{-\kappa L} \sim e^{-(|\gamma| + \kappa)L},
\end{equation}
\noindent where $1/\kappa=\xi_d$ is the localization length for the disorder
(in the real part) problem.  Thus the asymptotic behavior of the transmittance
for both the absorbing and the amplifying case is the same (i.e., exponential
decay) and the effective localization length $\xi$ for the disordered complex
site-potential case is given by $\xi^{-1}=l_a^{-1} + \xi_d^{-1}$.  This result
has been checked here by using a numerical transfer matrix method \cite{gs}
both for the amplifying and the absorbing case.  The differencei in the
behavior of $T$ between the two cases (sharp decay as opposed to an initial
rise with large oscillations) appears only upto $L \sim \xi$, and disappears
for $L \gg \xi$.  It was shown in I for the ordered case with $\xi_d=\infty$,
i.e., $\xi = l_a$.  This may be explained as follows.  For concreteness, let
us consider an $E$ close to zero (band centre).  Then $|d| > |c|$.  Now in the
absorbing case $\gamma >$ 0 making the exponentially increasing term in the
denominator dominating from the beginning and hence $T$ decays without much of
an intereference (and hence oscillation) from the beginning.  On the other
hand, in the amplifying case $\gamma <$ 0, and hence the exponentially growing
term in the denominator cannot dominate until some large enough length scale.
Also there is a lot of interference between the growing and the decaying terms
in the denominator in this case resulting into an initial rise in $T$ with
strong oscillations upto a crossover length $L_c=L_c(E, \eta)$ determined by
the approximate equality of the two terms in the denominator of Eq.(2), i.e.,
by $2|\gamma| L_c \sim$ ln$|d/c|$.
For the amplifying chain in the presence of disorder ($W > 0$),
another competition to the amplification peak ($T_{max}$) appears because of
localization effects and $L_c(E,W,\eta) \sim 1/2~\xi$ ln$|d/c| \to 0$ as
$\xi \to$ 0, i.e., either one or both of $W$ and $|\eta| \to \infty$.

It may be noted though that the reflectance or the two-probe resistance
$R=|r|^2$, where $r$ is the complex reflection amplitude (due to
backscattering) behaves differently from the transmittance.  The
asymptotically constant reflectance is reduced ($R_{\infty} <$ 1) for
$\eta >$ 0 (absorbing chain), and is amplified ($R_{\infty} >$
1) for $\eta <$ 0 (amplifying chain) by factors depending on the disorder
strength ($W$), $|\eta|$ and  $E$.  The behavior (on an average) is very
similar to that for the ordered case (see the figures in I).  For a fixed
$E$ and $\eta <$ 0, disorder not only reduces the peak position ($L_c$) as
discussed above, but also reduces the $T_{max}$ as well as the
asymptotically amplified value of $R_{\infty}$.  In Fig.1, we show this for
the case of $E=$0.1, $\eta=-$ 0.1 (the same as in the ordered case of I), and
for a disorder strength $W=$2.0.  One can check in this case that the
$l_a \simeq$ 20, and $\xi_d \simeq$ 25.  Thus the effective localization 
length should be $\xi = (1/l_a + 1/\xi_d)^{-1} \simeq$ 11, and this value
matches that obtained from Fig.1.  Further it will be noted that the peak
$T_{max}=$1.3 appears at $L_c=$12 (compared to the $T_{max} \simeq$ 2800
and $L_c=$ 68 $\simeq 3l_a$ in the ordered case), and the amplified
backscattering $R_{\infty}$= 6.3 (compared to 1600 in the ordered case). 

The issue of the phase distribution of the reflection amplitude $r$ is quite
important \cite{gj, frei3, mpl2} because the evolution of the
phase does have important bearings \cite{mpl2} on the evolution of the
resistance, these two quantities being coupled to each other.  In the
ordered case ($W$ = 0), the phase distribution is a $\delta$-function at an
angle dependent on $E$, $\eta$ (both magnitude and sign) and the length $L$
for a small system.  It is independent of $L$ and the sign of $\eta$ for
$L \gg l_a$ (see the expression for $r$ in I).  On the other hand for a
hermitean disorder case ($\eta$ = 0), the phase ($\phi$) distribution
$P_L(\phi)$ is typically quite non-uniform.  The distribution is uniform
only in the special case of weak localization ($L \sim \xi_d$).  It evolves
with length and approaches the stationary limit for $L \gg \xi_d$.
This stationary distribution $P(\phi)$ has in general two peaks whose
strengths increase (while the separation decreases) with $W$ \cite{mpl2}.
They approach a single, narrow peak only in the asymptotic limit of $W \to
\infty$.  Obviously in the presence of both disorder and non-hermiticity,
there should be a competition between the broadening effect on $P_L(\phi)$
due to a finite disorder and the $\delta$-function narrowing effect due to
the absorption/ amplification.  While it is generally accepted that a non-zero
$\eta$ suppreses phase fluctuation \cite{gj, frei3}, we find below that the
suppression even in the stationary limit ($L \gg \xi$) is only partial in the
sense that the phase fluctuation does not become a $\delta$-function unless
$|\eta|$ is very large.  Further we find that this stationary distribution is
independent of the sign of $\eta$ (whether amplifying or absorbing chain).
In this regard, we first show in Fig. 2 the evolution
of $P_L(\phi)$ towards its stationary distribution $P(\phi)$ for the case of
$E$=1.0, $W$=2.5, and $|\eta|$=0.2 (i.e., for an amplifying and an absorbing
chain).  For this choice, $l_a \simeq$ 10, and $\xi_d \simeq$ 12 (in lattice
units).  Thus the competition between the two effects should be quite strong.   
In Figs. 2(a) to 2(d), we show the evolution of $P_L(\phi)$ for $L$ = 3, 7, 20
and 200 respectively.  In all cases the histograms were obtained from 10,000
configurations.  In the Fig. 2(d), $L$ = 200 $\gg l_a$ or $\xi_d$, and hence
all the distributions have become stationary (no further change), and as one
may note there is no difference between the distributions for $\eta$= $+$0.2
or $-$0.2 in this limit.  Next in Figs. 3(a)-3(d), we show the stationary
distribution $P(\phi)$ as $|\eta|$ is increased from 0.5 to 3.0.  The quite
broad stationary distribution in Fig.2(d) for $\eta$ = 0.2 gets continually
shrunk at larger values of $|\eta|$.  But even for a large $|\eta|$ = 3.0 in
Fig.3(d) where the decay length scale $l_a \simeq$ 0.83 (less than a lattice
unit), the stationary distribution $P(\phi)$ is far from a $\delta$-function. 
In contrast to a recent work \cite{frei3}, the phase distribution is nowhere
even approximately Gaussian (particularly because of two peaks in general).
Further our results differ from another recent work \cite{gj} in that
$P(\phi)$ does not break apart into two narrow peaks as $|\eta| \to \infty$. 

Since the two-probe properties behave differently for absorbing and amplifying
chains and our main purpose here is to look for an unified description, we
study the evolution of the four-probe resistance $R_4=R/T$.  Another serious
reason for doing this is that the probability distribution of $T$ or $R$ for
the amplifying chains is not well-behaved in the sense that all the moments
of $T$ or $R$ (including the first one, e.g., $<T>$) diverge even for a finite
$L$ \cite{been, frei2}.  Since disorder plays an important role, we look at
the logarithmic resistance in the form of $u(L)$ = ln$(1+R_4)$.  We note that
if $Ab$ is the absorption coefficient then $R+T+Ab=1$, and if $Am$ is the
amplification coefficient then $R+T-1=Am$.  Thus while the quantity $u$ is
nothing but the negative of the logarithmic transmittance $-$ln$T$ in the
hermitean disorder case, $u$ = ln$(1-Ab)~-$ ln$T$ in the absorbing case and
$u$ = ln$(1+Am)~-$ ln$T$ in the amplifying case.  For our calculations, we
choose the same parameters as in Figs.2 and 3, and show in Fig.4 the evolutions
of $u(L)$ and $var(u) = <u^2> - <u>^2$ in double-logarithmic plots.  In the
Fig.4(a) for a hermitean case ($\eta$= 0), both the $<u>$ and $var(u)$ diverges
algebraically but with different exponents in the regime $0<L<L_2$ (where
$L_2 \sim \xi_d$) in accordance with a recent proposal by the author
\cite{mpl2} regarding the existence of a two-parameter scaling in that regime
and that in the strong localization limit $L \gg \xi_d$, they diverge with a
single exponent (unity) in accordance with the one-parameter scaling theory.
In Figs.4(b)-4(d), we introduce increasing amounts of non-hermiticity $\eta
= \pm$(0.2 - 3.0) and find some profound changes in the relative behavior of
the moments.  The first thing to note is that the moments of $u$ evolve with
$L$ almost identically for both the absorbing and the amplifying cases; the
difference cannot be shown in the scale of the figure.  This already gives
the hint that $u$ is the right variable to look at for an unified description
of both of these cases.  In contrast to the two-probe properties, these moments
of $u$ do not diverge and there are no odd-even (in $L$) oscillations.  Further
since there are two length scales involved in this problem, one sees a
crossover in Fig.4(b) for $|\eta|$ = 0.2 from absorption/ amplification
dominated (for $L \le l_a$ = 10 in this example) growth in $R_4$ to the growth
affected by localization effects as well (for $L \ge \xi_d$ = 12).  For larger
$|\eta|$ this crossover is still there at smaller lengths and may not be
clearly discernible.  Further, in contrast to the hermitean disorder case
where $var(u)$ crosses $<u>$ from the lower side (mildly fluctuating, weakly
localized) to the higher side (fluctuations dominated, strongly localized) in
the thermodynamic limit, the asymptotic $var(u)$ in the non-hermitean case is
always less than $<u>$.  Thus the fluctuations in $u$ are getting suppressed
by coherent amplification/ absorption and this may be interpreted as a
{\it destruction} of the effect of localization on $var(u)$ (but not on $<u>$).
Indeed for a large $|\eta|$, $var(u)$ may be
orders of magnitude smaller than $<u>$.  An example is the case of Fig.4(d) for
$|\eta|$= 3.0, where $var(u)$ is about 2.5 orders of magnitude smaller than 
$<u>$.  It clearly demonstrates how strongly the {\it coherent} amplification/
absorption works against the localization effects in destroying the
fluctuations in transport.  Yet at the same time, we must notice that this
destruction is never strong enough to bring back the self-averaging
character (or, diffusive behavior) to the resistance $R_4$.  Since this
destruction mechanism is an altogether coherent effect, it
cannot destroy the the exponential growth of the mean and the variance
as an incoherent effect (say, due to phonons) would have done.

To check if $P_L(u)$ is at least asymptotically normal (i.e., $P(R_4)$
log-normal), one may calculate the kurtosis parameter $K=1/2(3-c_4/c_2^2)$
where $c_2=var(u)$ and $c_4=(u-<u>)^4$ at each $L$.  For a normal distribution,
$c_4=3c_2^2$ and hence $K=$ 0.  In Fig.5, we plot $K(L)$ as a function of the
system size $L$ for the parameters of Fig.4(b).  It may not be surprising that
$K(L)$ is not close to zero ($P_L(u)$ far from normal) for small lengths, but
for much larger lengths $L \gg 10\xi$ (where $\xi \simeq$ 5), $K(L)$ still
fluctuates within [+0.1,$-$0.1].  Thus $u$ does not seem to be asymptotically
normal, and hence $R_4$ does not seem to be asymptotically log-normal. 
 
To summarize, we have looked at the effect of backscattering in a disordered
chain with coherently amplifying or absorbing site-potentials.  Disorder
enhances the exponential decay of transmittance in the presence of
absorbers, while it suppresses the transmittance peak (and the amplified
asymptotic reflectance) due to the amplifiers.  The stationary ($L \to \infty$)
probability density of the phase of the reflection amplitude is generally an
asymmetric double-peaked function (not even approximately Gaussian) whose
width is progressively suppressed towards zero ($\delta$-function-like form
as in the case of a pure amplifying/ absorbing chain) by larger amplifying/
absorbing strength $|\eta|$ or by larger disorder.  The four-probe resistance
$R_4$ gives an unifying description for both the amplifying and the
absorbing chain, and unlike in the case of Anderson localization the
amplifying/ absorbing effect can suppress the fluctuations of the logarithmic
resistance by orders of magnitude compared to the average logarithmic
resistance.  Study of the kurtosis indicates that the probability density
of the resistance $R_4$ is not asymptotically log-normal. 

\smallskip

The author acknowledges the support and the warm hospitality of the Condensed
Matter Research Group at the International Centre for Theoretical Physics,
Trieste, Italy where a substantial part of the work was done.

\newpage
{\bf Figure Captions:}\\

{\bf Fig.1} Evolution of the averages of the logarithmic reflectance ($<$ln$R>$)
and the logarithmic transmittance ($<$ln$T>$) as a function of chain length $L$,
for a coherently amplifying chain with $\eta=-$0.1,
Fermi energy $E$ = 0.1, and a disorder strength $W$ = 2.0.  The averages are
calculated at each $L$ for 10000 configurations.

{\bf Fig.2} Evolution of the distribution of the phase $P_L(\phi)$ with $L$
for both an amplifying and an absorbing chain $\eta=\pm$0.2, $E$ = 1.0, $W$ =
2.5.  Histograms representing the distribution are drawn using 10000
configurations each for (a) $L$ = 3, (b) $L$ = 7, (c) $L$ = 20, and
(d) $L$ = 200.  The distributions for the pure amplifying/ absorbing as well
as the hermitean ($\eta=$ 0) disorder cases are also shown for each $L$. 
The Fig.2(d) represents the stationary distribution $P(\phi)$.

{\bf Fig.3} Stationary phase distribution $P(\phi)$ for $E$ = 1.0 and $W$ = 2.5
using $L$ = 200 for (a) $|\eta|=$ 0.5, (b) $|\eta|=$ 1.0, (c) $|\eta|=$ 2.0 and
(d) $|\eta|=$ 3.0. 

{\bf Fig.4} Evolution of the mean and variance of $u=$ ln$(1+R_4)$, where $R_4$
is the four-probe resistance, as a function of $L$.  Both amplifying and
absorbing chains with 6000 configurations were used, no distinction between 
them can be made to the scale of the figure.  The parameters used are $E$ = 1.0,
$W$ = 2.5, and (a) $|\eta|=$ 0.0, (b) $|\eta|=$ 0.2, (c) $|\eta|=$ 1.0 and
(d) $|\eta|=$ 3.0. 

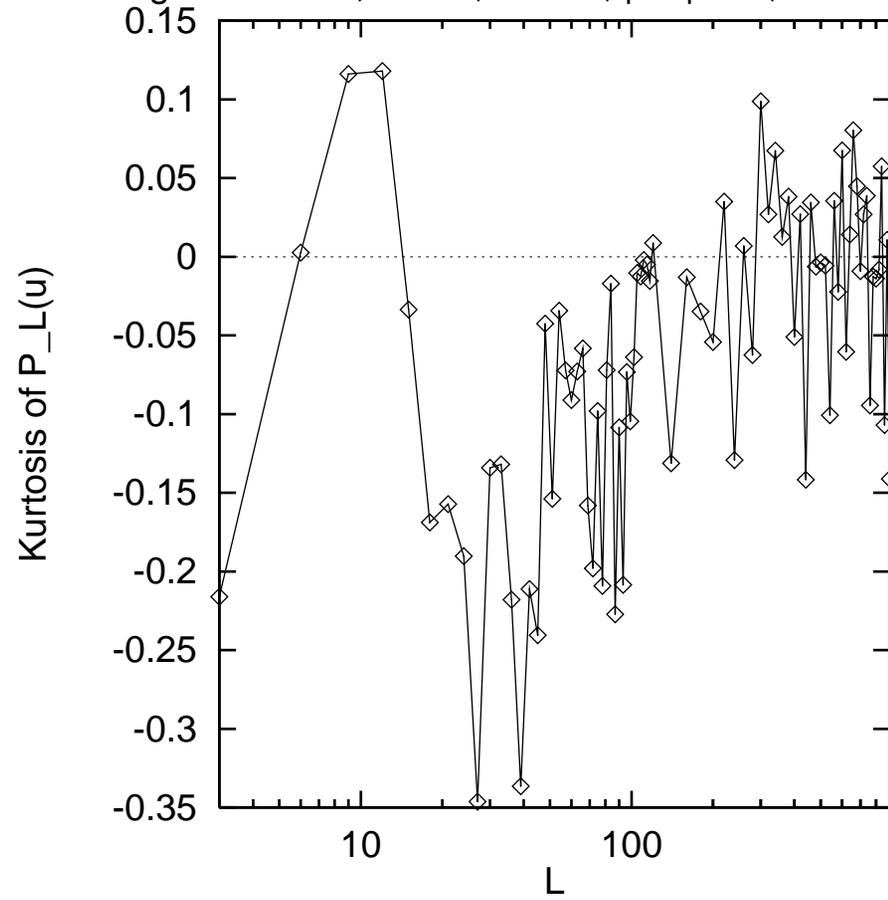
{\bf Fig.5} The dimensionless kurtosis (see text) parameter as a function of
length $L$ using $E$ = 1.0, $W$ = 2.5, $|\eta|$ = 0.2 and 6000 configurations.

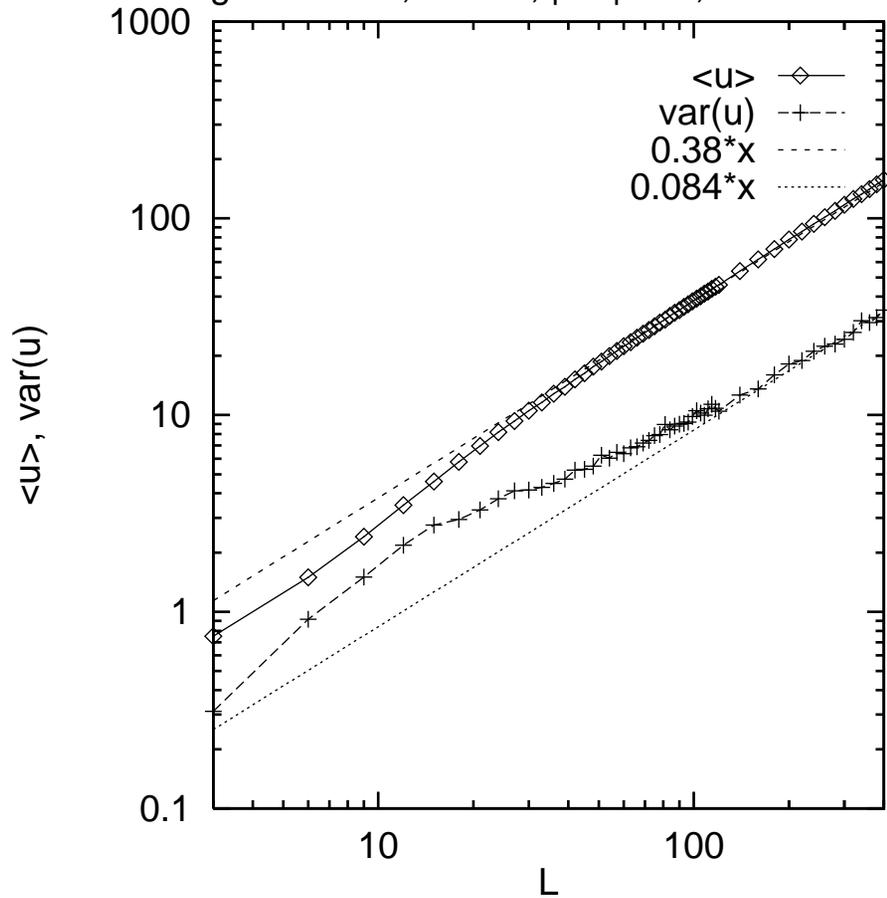
\end{document}